\documentclass[amsmath, amssymb, showkeys, titlepage, superscriptaddress, preprint]{revtex4-2}

\usepackage{float}
\usepackage{graphicx}
\usepackage{color}
\usepackage{tabularray}
\usepackage{multirow}
\usepackage{xr}
\usepackage[bookmarks=true, colorlinks=true, citecolor=black, linkcolor=black, urlcolor=black]{hyperref}
\usepackage{soul}
\usepackage{booktabs}
\begin{document}

\title{Periodical embeddings uncover hidden interdisciplinary patterns in the subject classification scheme of science}

\author{Zhuoqi Lyu}
\affiliation{Department of Data Science, College of Computing, City University of Hong Kong, Hong Kong, China}

\author{Qing Ke}
\email{q.ke@cityu.edu.hk}
\affiliation{Department of Data Science, College of Computing, City University of Hong Kong, Hong Kong, China}

\date{\today}

\begin{abstract}
Subject classification schemes are foundational to the organization, evaluation, and navigation of scientific knowledge. While expert-curated systems like Scopus provide widely used taxonomies, they often suffer from coarse granularity, subjectivity, and limited adaptability to emerging interdisciplinary fields. Data-driven alternatives based on citation networks show promise but lack rigorous, external validation against the semantic content of scientific literature. Here, we propose a novel quantitative framework that leverages classification tasks to evaluate the effectiveness of journal classification schemes. Using over 23 million paper abstracts, we demonstrate that labels derived from $k$-means clustering on Periodical2Vec (P2V)---a periodical embedding learned from paper-level citations---yield significantly higher classification performance than both Scopus and other data-driven baselines (e.g., citation, co-citation, and Node2Vec variants). By comparing journal partitions across classification schemes, two structural patterns emerges on the map of science: (1) the reorganization of disciplinary boundaries---splitting overly broad categories (e.g., “Medicine” into ``Oncology'', ``Cardiology'', and other specialties) while merging artificially fragmented ones (e.g., “Chemistry” and “Chemical Engineering”); and (2) the identification of coherent interdisciplinary clusters---such as ``Biomedical Engineering'', ``Medical Ethics'', and ``Information Management''---that are dispersed across multiple categories but unified in citation space. These findings underscore that citation-derived periodical embeddings not only outperform traditional taxonomies in predictive validity but also offer a dynamic, fine-grained map of science that better reflects both the specialization and interdisciplinarity inherent in contemporary research.

\end{abstract}


\maketitle

\section{Introduction}

Scientific knowledge is organized through subject classification systems that serve as fundamental infrastructures in the scholarly ecosystem. These systems play crucial roles across the research lifecycle: authors rely on these systems when submitting manuscripts to indicate relevant research areas, while readers use them to navigate the vast ocean of publications, and provide the essential framework for scientometric analyses that map the landscape of science~\cite{glanzel2003new, waltman2012new}. The significance of these classifications extends beyond mere categorization---they fundamentally shape how scientific findings are communicated, evaluated, and ultimately how the evolution of science is measured and guided. Beyond academic functions, they profoundly affect information acquisition and management, individual/institutional academic performance evaluation~\cite{zhang2025impact}, research grant allocation, and science policy making~\cite{rafols2009content}. 

Traditional classification schemes, such as those provided by Web of Science (WoS) and Scopus, which are mostly designed by in-house experts manually, assign hard-coded categorical labels to papers, journals, conference proceedings, and books. Despite the widespread use and importance of these classification schemes, they often suffer from limitations such as lack of transparency in their methodology~\cite{wang2016large}, subjectivity~\cite{pudovkin2002algorithmic}, lack of granularity~\cite{leydesdorff2016operationalization, zhang2025impact}, limited coverage, and the inability to adapt quickly to emerging interdisciplinary fields~\cite{thelwall2024accuracy}, hence may not fully capture the nuanced and evolving nature of scientific research, leading to potential misclassifications or oversights~\cite{shen2024under}, and unfair publication evaluation~\cite{liao2025journal}. 

Several attempts have been made to improve upon or replace the WoS ~\cite{janssens2009hybrid, thijs2015bibliographic, zhang2010journal} and Scopus ~\cite{gomez2011improving, gomez2014optimizing, gomez2016updating} classification schemes. These data-driven methods primarily mine the nuanced citation relationships between papers, offering better scalability and flexibility compared to traditional methods. However, despite these attempts, two critical gaps remain. First, there is a lack of systematic validation using external data sources beyond the citation relationship itself ~\cite{wang2016large}, to cross-verify the rationality of the proposed classifications. Rich identity information embedded in text, keywords, or grant categories is not yet fully explored and utilized for testing purposes. Second, existing evaluations primarily rely on self-referential metrics like clustering coherence and concentration ~\cite{gomez2016updating}, or manual inspection using author-defined thresholds ~\cite{wang2016large}, without establishing universal quantitative measures that would enable objective comparisons across different classification schemes. These limitations highlight the need for both rigorous cross-modal validation benchmarks and evaluation metrics to assess the effectiveness of different classification schemes against extensive academic periodicals. 


To address these challenges, this study aims to apply advanced embedding methods to large-scale citation networks, thereby generating more meaningful data-driven subject assignments for journals. Following our past research ~\cite{peng2021neural, lyu2025mapping}, we sample the paper-paper citation network to generate 107,107,571 citation trails and apply Word2Vec ~\cite{tomas2013sgns} on them, seeking to generate vector representations of periodicals that reflect their semantic relationships. Then, classification labels can be obtained by applying clustering algorithms on embeddings. We then evaluate the effectiveness of these embeddings in comparison to Scopus and other data-driven classification shcemes, through the lens of semantic-based classification tasks and various metrics of topic consistency. Specifically, we first perform document classification on over 23 million paper abstracts to assess validity of proposed disciplinary partition for scientific periodicals. Second, we conduct topic modeling analysis to measure the alignment between the natural topical structure in research papers and various classification schemes. This controlled methodology enables us to objectively evaluate how well different approaches align with the actual organization and structure of the scientific ecosystem, as reflected in the content of research papers. Furthermore, we find that periodical embedding can not only enhance the accuracy of subject classification but also provide insights into the evolving landscape of scientific publishing. By identifying hidden interdisciplinary connections and emerging research areas, it contributes to a more nuanced understanding of scientific progress and its implications for academic evaluation, research funding, and policy making. 

\section{An overview of journal classification systems}

\subsection{Traditional expert-based classification systems}

The most widely used journal subject classification systems are those developed and maintained by large bibliographic databases like Web of Science (WoS) and Scopus. WoS currently categorizes journals into 254 subject categories, while Scopus employs the All Science Journal Classification (ASJC) system, which comprises 27 primary subject areas. These schemes are mainly curated by in-house experts who rely on editorial judgment, journal scope statements, and, in some cases, citation patterns to assign subject categories to journals~\cite{glanzel2003new, pudovkin2002algorithmic}. Such expert-curated classification frameworks play a central role in bibliometrics and research evaluation, supporting applications ranging from the calculation of field-normalized citation metrics~\cite{leydesdorff2016operationalization} to research assessment exercises.

However, expert-based classification systems have several well-recognized limitations. First, the assignment process is often opaque and subjective, leading to inconsistencies across databases and over time~\cite{pudovkin2002algorithmic, leydesdorff2016operationalization}. Second, the granularity of categories is uneven: certain biological science fields are divided into many specialized categories, whereas disciplines like medicine are under-partitioned and grouped together as broad categories~\cite{zhang2010journal, leydesdorff2016operationalization}. Third, both WoS and Scopus's classification systems focus almost exclusively on journals, with limited or no coverage of conference proceedings, which are the dominant publication outlets in certain rapidly advancing fields like computer science. This limited coverage further restricts their ability to reflect the full spectrum of scientific activities and interdisciplinary connections. Lastly, the labor-intensive nature of expert curation impedes scalability and limits the ability of these systems to adapt promptly to shifts in the scientific landscape. 

\subsection{Algorithmic and data-driven approaches}

In addition to expert-based classifications, researchers have developed a range of algorithmic and data-driven alternatives. The mainstream of studies leverages the structure of citation networks to depict the organization of science and generate new classification schemes. For example, \cite{zhang2010journal} employed cross-citation analysis of journals, combined with hierarchical clustering, to validate and refine subject classifications. This approach demonstrated that data-driven methods could identify misclassified journals and reveal latent disciplinary structures more objectively than expert curation. Similarly, using established journal categories as reference, \cite{boyack2005mapping} systematically compared five science maps based on citation and co-citation similarity measures, highlighting the potential of network-based methods for revealing scientific structure. Subsequent research has employed more sophisticated algorithms to optimize journal classification via citation flows~\cite{gomez2011improving, gomez2014optimizing, gomez2016updating}. It has also developed hybrid methods that combine citation data, bibliographic coupling, and textual analysis to improve classification robustness and specificity~\cite{janssens2009hybrid, thijs2015bibliographic, gomez2016updating}. More recently, advances in graph embedding techniques have enabled the application of representation learning to scientific networks~\cite{peng2021neural, lyu2025mapping}, showing potential to yield fine-grained, scalable, and adaptive classification systems that better capture the inter-connectivity of the evolving scientific communication. These developments reflect a growing consensus that data-driven, algorithmic frameworks can not only complement and, in some cases, surpass expert-based systems in mapping the evolving structure of science.

Despite these methodological innovations, there is still a lack of external validation and benchmarking of various journal classification schemes against a vast volume of research papers. Most existing evaluations rely on manual inspection of category assignments~\cite{pudovkin2002algorithmic, gomez2016updating, leydesdorff2016operationalization}, or internal clustering metrics like modularity~\cite{janssens2009hybrid}, silhouette score~\cite{janssens2009hybrid, thijs2015bibliographic} and cosine similarity~\cite{rafols2009content}. Systematic cross-modal validations using non-citation data sources---such as text content, author-supplied keywords, or funding agency categories---remain rare~\cite{wang2016large, thelwall2024accuracy}. The lack of widely accepted external benchmarks hinders quantitative and objective comparisons between classification schemes, limiting our ability to assess their semantic coherence, practical utility, and adaptability to new or interdisciplinary fields.

\section{Results}

\subsection{Document classification task}
\label{sec:doc-class}

To systematically and quantitatively evaluate the effectiveness of different periodical classification schemes $\mathcal{C}$, we conduct a large-scale document classification task using $23,322,430$ paper abstracts from the Microsoft Academic Graph (MAG) dataset. The fundamental principle guiding our design is that a good classification scheme should $\mathcal{C}_i$ enable a trained classifier to effectively predict the subject areas of papers based on their semantic content. If $\mathcal{C}_i$ truly captures meaningful disciplinary boundaries, then papers assigned to the same category should exhibit semantic similarity, which should be detectable by computational methods. More specifically, given the same training corpora, we expect that a more meaningful partition of periodicals would result in a classifier with a better capability to retrieve papers from the same category, that is, a higher \emph{recall}. By holding identical input textual features and model structure, while varying only the labels derived from different $\mathcal{C}_i$, we isolate the effect of the classification schemes themselves.

Here, we compare six distinct classification schemes $\mathcal{C}$: (1) the conventional Scopus subject classification; $k$-means applied to (2) Citation Matrix; (3) Citation+Node2Vec, (4) Co-citation matrix, (5) Co-citation+Node2Vec, and (6) Periodical2Vec. For each $\mathcal{C}_i$, periodical cluster labels (or ASJC subject areas for Scopus, see \S~\ref{subsection:Data}) are used as the target classes, and an identical combination of Hashing Vectorization and Complementary Naive Bayes (CNB) classifier was employed for model training and evaluation. The classification results were used to evaluate the effectiveness of $\mathcal{C}_i$ in generating meaningful groupings of periodicals. Higher scores indicate that the partitions obtained through a scheme are more consistent with the semantic information embedded in the abstracts of papers from the MAG dataset.

We set up two prediction scenarios: multi-class classification and one-vs-rest classification. In the former, to evaluate each $\mathcal{C}_i$, we use a single classifier to predict the class label of a paper. Table~\ref{tab:multi_class_results_transposed} summarizes the performances of the six methods in terms of macro and weighted averages for precision, recall, and F1-score. Macro values here refer to the unweighted average of scores over classes. While macro precisions are similar, the partition based on Periodical2Vec (P2V) results in significantly higher recall, meeting our expectations. Node2vec, which aims to learn a dense vector from the vanilla sparse citation vector, is unable to improve. Note that the distribution of periodicals across categories is highly imbalanced in $\mathcal{C}^{Scopus}$ ( Table~S1), with a large proportion of journals belonging to the field of \emph{Medicine}, due to the coarse granularity of Scopus (see \S~\ref{subsec:structure}). This inherently provides $\mathcal{C}^{Scopus}$ with a natural advantage on weighted metrics, as the performance on the largest categories dominates the weighted averages. Despite this advantage for Scopus, our citation-embedding-based scheme ($\mathcal{C}^{P2V}$) consistently outperforms all alternatives, across both macro and weighted metrics. Specifically, our method achieves the highest recall and F1-score, indicating robust performance across all classes, including those with fewer samples. This demonstrates that our method provides substantial gains in overall classification consistency and generalizability.

\begin{table}[t!]
\centering
\caption{Test set multi-class classification performance comparison between classifiers trained using different labels. Results are averaged over 10-fold cross-validation; values are reported as mean~$\pm$~standard deviation.}
\label{tab:multi_class_results_transposed}
\tiny
\begin{tabular}{l|ccc|ccc|cc}
\toprule
 & \multicolumn{3}{c|}{Macro} & \multicolumn{3}{c|}{Weighted} & \multicolumn{2}{c}{Rank}\\
 & Precision & Recall & F1 & Precision & Recall & F1 & Avg Precision & Loss \\
\midrule
Periodical2Vec & $60.62\pm 0.24$ & $\textbf{50.62}\pm 0.04$ & $\textbf{51.69}\pm 0.04$ & $61.21\pm 0.10$ & $60.38\pm 0.03$ & $\textbf{58.81}\pm 0.03$ & $\textbf{74.47}\pm 0.02$ & $\textbf{0.0550}\pm 0.0001$ \\
Citation & $60.70\pm 0.59$ & $41.67\pm 0.03$ & $44.26\pm 0.04$ & $58.54\pm 0.03$ & $57.87\pm 0.03$ & $56.91\pm 0.03$ & $72.17\pm 0.02$ & $0.0853\pm 0.0001$ \\
Citation+Node2Vec & $46.31\pm 0.27$ & $33.22\pm 0.03$ & $31.17\pm 0.04$ & $47.13\pm 0.08$ & $47.78\pm 0.03$ & $41.43\pm 0.03$ & $64.59\pm .02$ & $0.0828\pm 0.0001$ \\
Co-citation & $\textbf{62.93}\pm 0.17$ & $47.50\pm 0.04$ & $49.14\pm 0.05$ & $61.14\pm 0.03$ & $60.40\pm 0.02$ & $58.76\pm 0.02$ & $74.45\pm 0.02$ & $0.0598\pm 0.0001$ \\
Co-citation+Node2Vec & $38.57\pm 2.09$ & $12.51\pm 0.03$ & $12.75\pm 0.05$ & $59.52\pm 0.32$ & $\textbf{60.86}\pm 0.03$ & $57.29\pm .03$ & $74.42\pm 0.02$ & $0.1294\pm 0.0002$\\
Scopus & $60.93\pm 0.38$ & $34.41\pm 0.03$ & $36.28\pm 0.03$ & $\textbf{61.35}\pm .07$ & $60.20\pm 0.04$ & $55.90\pm 0.04$ & $74.07\pm .03$ & $0.0574\pm 0.0001$ \\
\bottomrule
\end{tabular}
\end{table}

\begin{figure*}[]
\centering
\includegraphics[width=\linewidth]{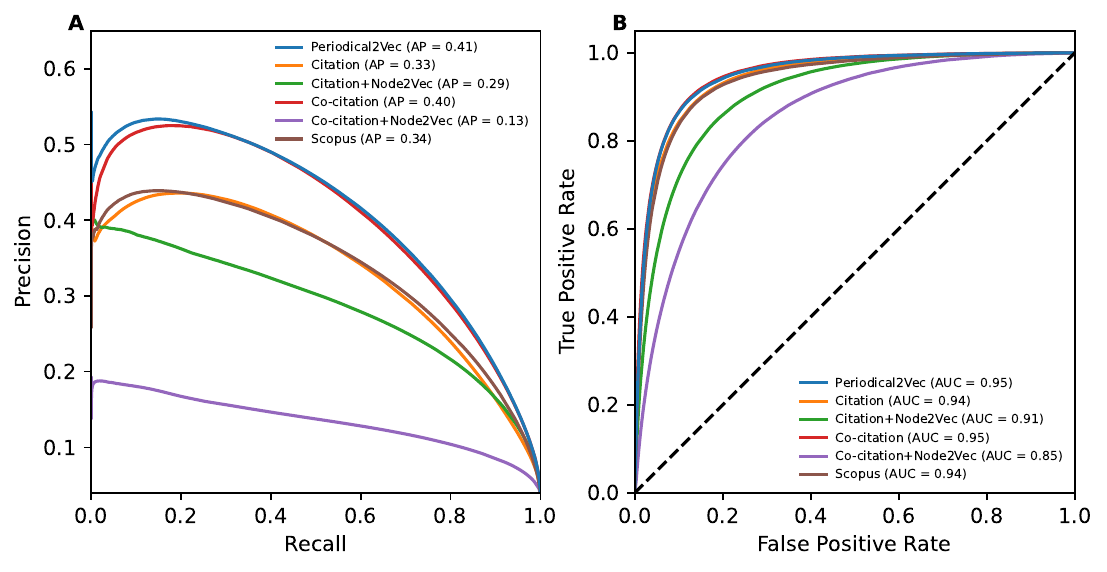}
\caption{One-vs-rest classification performance comparison between classifiers trained using different classification labels. Curves are plotted as the macro average of all individual classes' results.
\textbf{(A)} Precision-Recall curves. The number in parentheses indicates average precision. 
\textbf{(B)} Receiver operating characteristic curves. The number in parentheses indicates the area under the curve.
}
\label{fig:curves}
\end{figure*}

For the one-vs-rest classification scenario, we iteratively train $K$ different binary classifiers where $K$ is the number of total classes. For each classifier, we choose one class as positive, while all other classes are negative. Fig.~\ref{fig:curves} presents the macro-averaged Precision-Recall (PR) and Receiver Operating Characteristic (ROC) curves for each approach. $\mathcal{C}^{P2V}$ achieves the best PR performance and the highest ROC area under the curve (AUC), outperforming all other alternatives. We emphasize that in highly imbalanced binary-class settings like ours, metrics based on PR curves are more informative and reliable than ROC AUC for measuring the real-world effectiveness of classifiers~\cite {davis2006relationship}. The superior macro-average AP of our method, therefore, demonstrates not just strong discrimination overall, but especially improved performance in retrieving relevant items from underrepresented subject areas.

Taken together, these results confirm the superior discriminative power and class separability of the labels derived from our periodical embeddings, both in general multi-class settings and in distinguishing individual classes from the rest.

\subsection{Topic-label consistency}
\label{sec:lda}

To further assess the semantic coherence and validity of different classification schemes, we conduct an agreement analysis between topics detected from papers' abstracts and labels based on classifications. This analysis aims to examine whether the labels align with the natural topic distribution in the academic literature. Specifically, we employ Latent Dirichlet Allocation (LDA)~\cite{blei2003latent} to the entire corpus of abstracts, thereby inferring a set of latent topics that capture the underlying thematic structure of the scientific literature. Each document is assigned to its most probable topic, and these topic assignments were then compared to the periodical labels generated by each classification approach. The alignment between LDA-inferred topics and classification labels is quantitatively evaluated using three widely adopted clustering agreement metrics, namely Normalized Mutual Information (NMI), Adjusted Rand Index (ARI), and Fowlkes-Mallows Index (FMI). 

Table~\ref{tab:topic_label_consistency} presents the results, $\mathcal{C}^{P2V}$ achieves the highest scores across all three metrics (NMI = 0.3195, ARI = 0.1693, FMI = 0.2169), followed by the $\mathcal{C}^{Co-citation}$. Both $\mathcal{C}^{N2V}$ and $\mathcal{C}^{Citation}$ yield lower agreement with the LDA-inferred topical structure, compared to Scopus. These results indicate that the clusters derived from citation-based periodical embeddings most closely align with the underlying topical structures of scientific literature, outperforming both alternative data-driven methods directly derived from raw citations and the established expert-curated Scopus classification.

\begin{table}[h!]
\centering
\caption{Normalized Mutual Information, Adjust Rand Index, and Fowlkes–Mallows Index between papers' topic assigned by LDA and label assigned by different classification schemes. }
\label{tab:topic_label_consistency}
\scriptsize
\begin{tabular}{l|c|c|c|c|c|c} 
\toprule
& {Periodical2Vec} & {Citation} & {Citation+Node2Vec} & {Co-citation} & {Co-citation+Node2Vec} & {Scopus} \\ 
\midrule
NMI & \textbf{0.3195} & 0.2662 & 0.2748 & 0.3168 & 0.1672 & 0.3010\\
ARI & \textbf{0.1693} & 0.1127 & 0.1306 & 0.1633 & 0.0324 & 0.1371\\
FMI & 0.2169 & 0.1804 & 0.1816 & \textbf{0.2203} & 0.1631 & 0.2106\\ 
\bottomrule
\end{tabular}
\end{table}

Collectively, these findings provide robust quantitative evidence that the citation-embedding-based classification method not only enhances the performance of downstream tasks like document classification but also yields label assignments that are more semantically coherent with the intrinsic topical structure of scientific research.

\subsection{Structure of Science} 
\label{subsec:structure}

The preceding analyses have systematically validated the effectiveness of our citation-based classification system through quantitative evaluations. Building upon these foundations, we now shift our focus to more exploratory investigations—examining how journals are assigned differently across classification schemes and uncovering hidden patterns that reflect the evolving structure of scientific research.

\begin{figure}[t!]
\centering
\includegraphics[width=1\linewidth]{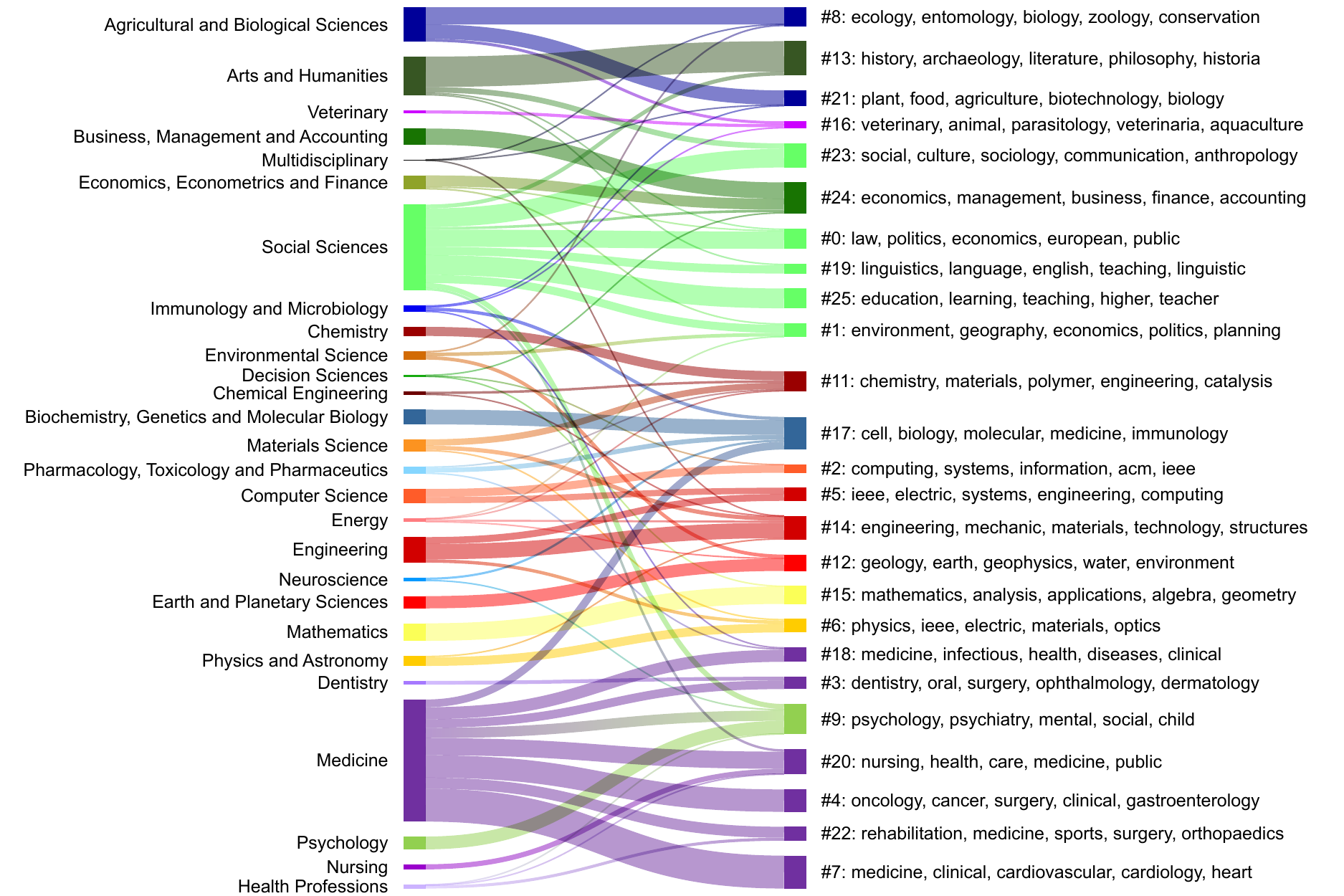}
\caption{A Sankey diagram visualizing how journals are assigned differently between the Scopus categories (left-side nodes) and the data-driven clusters (right-side nodes). The top 5 frequent words from journal names are presented for each node on the right side to indicate cluster topics. The thickness of flows quantifies the proportion of journals redistributed between the two classification schemes. For visual clarity, we filter out flows with less than $\min(10\%\times N_{source},50)$ journals, where $N_{source}$ is the number of journals in source nodes (the left side). An interactive online version of this Sankey diagram can be accessed at \url{https://lyuzhuoqi.github.io/periodical-clustering/sankey/snakey_kmeans_filtered.html}.}
\label{fig:sankey}
\end{figure}

Fig.~\ref{fig:sankey} uses a Sankey flow visualization to examine how journals are assigned differently between Scopus's classification and our data-driven classification. First, it indicates that the distribution of data-driven clusters based on P2V is more balanced than the Scopus classification scheme; while 23\% and 16\% of journals in $\mathcal{C}^{Scopus}$ are categorized as ``Medicine'' and ``Social Science'', the largest cluster in $\mathcal{C}^{P2V}$ contains 6\% journals. Underlying the differing size balance is both alignment and divergence between the two classification schemes. Specifically, the data-driven clustering reveals finer substructures within Scopus's board categories. For example, journals classified under ``Medicine'' in Scopus are distributed across multiple clusters, like cluster 7, which concentrates on cardiology, cluster 4, which focuses on oncology, and several others---reflecting the rich internal structure that is collapsed into a single category by Scopus. Likewise, Social Sciences journals are split into several specialized clusters like cluster 19 ``Linguistics'' (e.g., \emph{Language and Literature}) and \#25 ``Education'' (e.g., \emph{Teaching and Teacher Education}), diverging from Scopus's monolithic grouping. These patterns suggest that traditional classifications often conflate fields with distinct citation exchanges. 

Meanwhile, some $\mathcal{C}^{Scopus}$ categories like Mathematics have a one-to-one correspondence in $\mathcal{C}^{P2V}$, and some others are combined into the same clusters. For instance, most Chemical Engineering and Chemistry journals are assigned to cluster 11; similarly, most journals labeled as ``Economics, Econometrics and Finance'' and ``Business, Management and Accounting'' are assigned to cluster 24. Both cases reflect a potential over-partition drawback of Scopus classification, that is, these fields are cohesively grouped into the same cluster through citation patterns inherently intertwined in research practices, while they are artificially separated by Scopus. Apart from amalgamation, small categories, including ``Dentistry'', ``Veterinary'', ``Psychology'', and ``Earth and Planetary Sciences'', which are predominantly mapped to cluster 3, 16, 9, and 12, expanded due to the intake of flow from other Scopus categories, highlighting how citation-based embeddings capture interdisciplinary connections and subfield granularity. Through the consolidation of fragmented Scopus categories and the decomposition of oversized categories, a natural size equilibrium can be achieved that reflects actual research activity scales.

\begin{figure}[t!]
\centering
\includegraphics[width=0.9\linewidth]{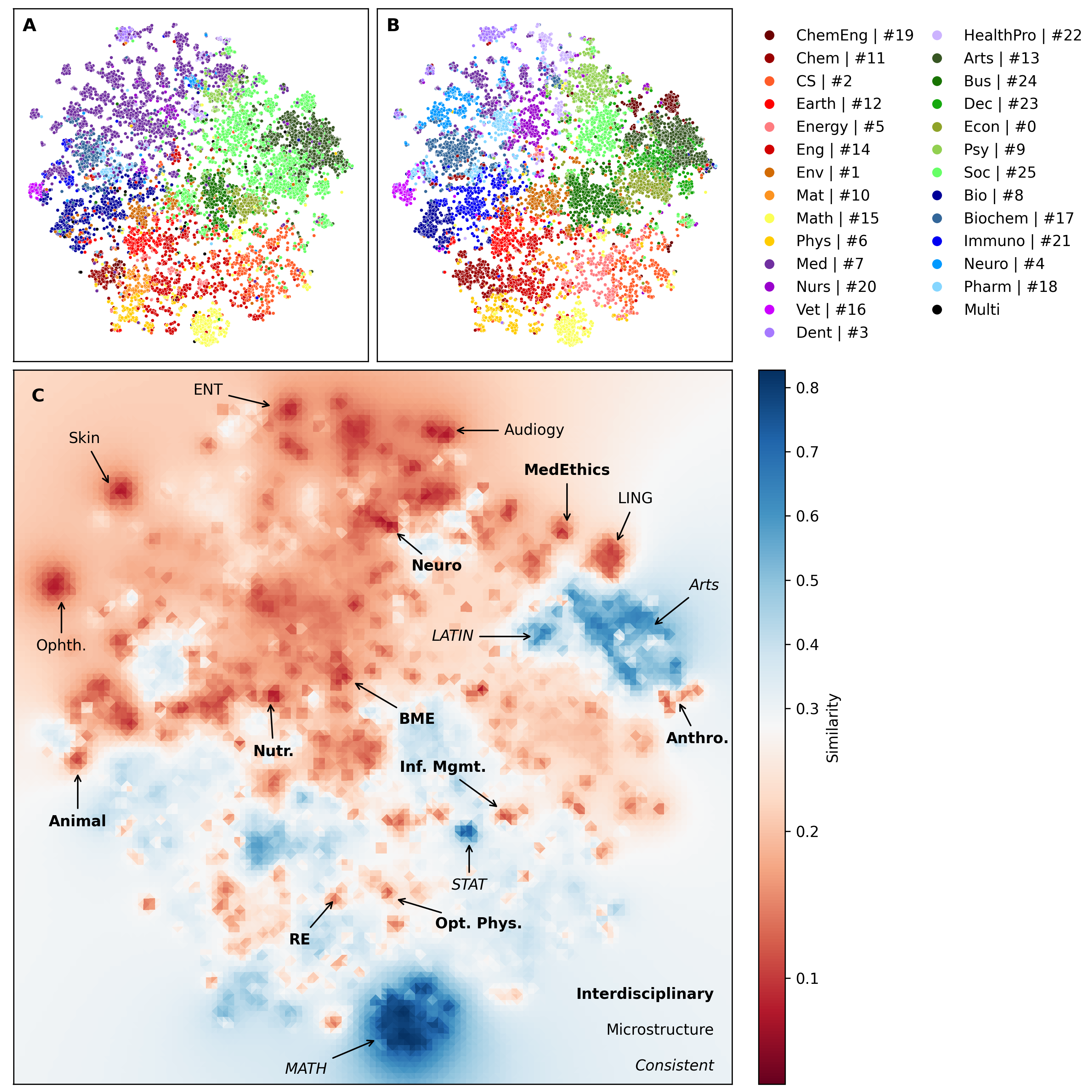}
\caption{Comparing $\mathcal{C}^{Scopus}$ and $\mathcal{C}^{P2V}$. 
(A--B) 2-D projections of journal P2V embeddings using t-SNE~\cite{van2008visualizing}. Legends are shared between (A) and (B) to enable visual cluster tracking. 
\textbf{(A)} Partitioned by $\mathcal{C}^{Scopus}$. Color denotes label. 
\textbf{(B)} Partitioned by $\mathcal{C}^{P2V}$. Color denotes cluster \#. 
\textbf{(C)} A interpolated heatmap of element-wise clustering similarity between Scopus and clustering-generated label, with exemplary topic-cohesive clusters which have notable low/high similarity annotated (check detailed cluster members in Tables~S2--S5). The line types of arrows indicate the reason for the unusual similarity between 2 classification schemes. The interpolation is based on inverse distance weighting (IDW) using power parameter=2.
}
\label{fig:scimap}
\end{figure}

To further map the nuanced structure of science, in Figs.~\ref{fig:scimap}A--B we show 2-D projections of journals based on their P2V emebddings and color them based on labels assigned in $\mathcal{C}^{Scopus}$ and $\mathcal{C}^{\text{P2V}}$ respectively, allowing us to observe how journals are distributed and interacted, furthermore, microscope the cluster boundaries in details. Fig.~\ref{fig:scimap}A shows that numerous journals near cluster peripheries receive conflicting Scopus labels (e.g., journals lying on the edge of a cluster of ``Veterinary'' are labeled as ``Agricultural and Biological Sciences''). In contrast, the $k$-means clusters exhibit crisper boundaries (Fig.~\ref{fig:scimap}B). 

To quantify this structural disparity, we calculate the element-wise clustering similarity~\cite{gates2019element}---a sample-level metric comparing label consistency across classification systems, and then apply spatial interpolation using Inverse Distance Weighting~\cite{shepard1968two} (IDW), to generate Fig.~\ref{fig:scimap}C. The visualization highlights both the agreement and disagreement between manual classification and actual citation patterns. For instance, we annotate regions with high agreement using italic text in Fig.~\ref{fig:scimap}C, including ``Mathematics'', ``Statistics'', ``Arts, Latin American studies'', etc. For regions with high disagreement (low similarity), we observe two distinct patterns. The first corresponds to intricate substructures within broad disciplinary categories, including ``Otolaryngology'', ``Ophthalmology'', and ``Dermatology''. These domains exhibit particularly fragmented microstructure patterns (annotated using normal text in Fig.~\ref{fig:scimap}C), aligning with what we observe in Fig.~\ref{fig:sankey}, indicating an artificial conflation of specialized fields in Scopus, though they show distinct citation patterns from their mainstream medicine peers. The second case of high disagreement corresponds to topic-cohesive journal clusters that are characterized by high interdisciplinarity (e.g., ``Information Management'', ``Biomedical Engineering'', and ``Medical Ethics'', annotated with bold text in Fig.~\ref{fig:scimap}C), form volatile heterogeneous clusters under either or both of $\mathcal{C}^{Scopus}$ and $\mathcal{C}^{\text{P2V}}$, which reveals their interdisciplinary nature. A detailed list of periodical members located at these fluid boundaries is presented in Tables~S2--S3.

Overall, the persistent alignment in certain domains confirms that citation networks can capture essential features of scientific practice, while the observed divergences highlight the dynamic, interdisciplinary, and fine-grained nature of contemporary scientific research.

\section{Discussion}

The main contribution of this work is to introduce the utilization of prediction tasks for the quantitative evaluation of different journal classifications. Through semantical cross-validation based on paper abstracts, we demonstrate that our embedding-based method not only aligns better with paper-level textual information but also yields better performance in downstream tasks, compared to other citation-based methods as well as expert-curated taxonomies like Scopus. This capability positions our method as a powerful tool for depicting and microscoping the evolving landscape of science. By revealing hidden interdisciplinary connections, consolidating fragmented fields, and exposing substructures within broad categories, these data-driven embeddings provide a dynamic representation of scientific knowledge that better aligns with actual research practices. 




We acknowledge two main limitations in this work. First, the selection of journal classification systems for comparison is not exhaustive. For instance, other widely used schemes like WoS subject categories are not covered here but may be examined in the future. Second, the single-vector representation may inadequately capture the contextual diversity of multidisciplinary periodicals, as seen in cases like \emph{PNAS} where a journal's scope spans multiple distinct research communities. Third, despite our citation-based approach outperforming its counterparts, we recognize untapped potential in combining citation patterns with textual semantics. Recent advances in multimodal learning\cite{xu2023multimodal} suggest that fusing citation embeddings with text-based representations~\cite{beltagy2019scibert, dong2024towards} might be able to better capture the topology of science entities. Despite these constraints, our framework establishes a foundation for several promising directions. Future research could explore: (1) hybrid models that integrate citation embeddings with state-of-the-art NLP representations (e.g., transformer-based embeddings), moving beyond bare metadata to contextual language understanding; (2) how such multimodal systems could enhance emerging-field detection, particularly in fast-moving domains where innovation embedded in text precedes citation burst.

\section{Data and Methods}

\subsection{Data} \label{subsection:Data}

In this work, we use a version of the Microsoft Academic Graph (MAG) retrieved in December 2021~\cite{sinha2015overview}, which contains $96,072,424$ papers published in $49,006$ journals and $4,551$ conferences between 1800 and 2021. As the landscape of science evolves fast over time~\cite{lyu2025mapping}, we restrict our analysis to journal and conference papers published from 2010 onward, to focus on the latest snapshot. To avoid bias toward history, we only retain papers published in the corresponding decade, along with citation relationships between them (i.e., both the source and terminal publications in a single citation record must have been published in the same decade). Finally, our dataset contains $41,820,928$ papers and $359,442,336$ citations from $46,074$ periodicals. 

We used Scopus subject area categories to label journals, as a manual-annotation counterpart of our self-annotation labels, for benchmarking. Scopus is a database that covers 40,878 peer-reviewed journals in top-level subject fields: life sciences, social sciences, physical sciences, and health sciences~\cite{elsevier-no-date-scopus-content}. It used the ASJC (All Science Journal Classification) scheme to categorize journals into 27 major subject areas (including ``multidisciplinary''), and they may belong to more than one area simultaneously. We assign the subject area, shared by most of a specific journal's top 50 similar peers (depending on cosine similarities between journals' P2V embeddings), to this journal as its mono-label. 

We finally matched $20,038$ journals between our embeddings and Scopus based on journal names. It is worth noting that labels of conference proceedings are not provided by Scopus, while those conferences are also included during the construction of embeddings.

\subsection{Classification Schemes}

We implement five algorithmic methods to get journal classifications, which all involve applying $k$-means to periodical vectors. The first is the P2V, which was previously prepared for $37,594$ periodicals~\cite{lyu2025mapping}. 

Then, for the other four alternative vector representations (Citation, Citation+Node2Vec, Co-citation, and Co-citation+Node2Vec), we construct a $43982 \times 43982$ periodical-periodical citation matrix, where each entry $E_{ij}$ records the number of citations from periodical $i$ to periodical $j$. Specifically, for each paper that cites a paper published in periodical $j$, the citation from the citing periodical $i$ to the cited periodical $j$ is counted. This matrix captures the citation flow between journals, where the value $E_{ij}$ corresponds to the total number of citations made by periodicals in journal $i$ to papers published in journal $j$. We normalize the matrix by row; therefore, each entry in row $i$ indicates the possibility of the journal $i$ citing other journals. 

The second vector representation (Citation) is to directly use the rows of this matrix as high-dimensional vectors for each periodical, where the raw citation information between journals is encoded in the matrix.

The third (Citation+Node2Vec) constructs a periodical--periodical citation network based on the unnormalized citation matrix. In this network, nodes represent periodicals, and edges represent the citation flow between them. We then apply Node2Vec~\cite{grover2016node2vec}, a commonly used graph embedding technique, on this citation network to obtain vector representations of periodicals. We use SNAP implementation~\cite{leskovecSNAPGeneralPurposeNetwork2016} and follow the default hyperparameters, with embedding dimension $d=128$, walk length $l=80$, number of walks per source $r=10$, return hyperparameter $p=1$, and inout hyperparameter $q=1$. The key difference between this method and our approach is that while we perform random walks on the paper-level citation network (where each paper is a node), this method works on the periodical-level citation network, which may result in the preservation of more fine-grained information in our approach.

The fourth (Co-citation) constructs a $43728\times 43728$ periodical-periodical co-citation matrix, where each entry $E_{ij}$ records the number of times periodical $i$ and periodical $j$ are co-cited by the same paper. More specifically, for each paper that cites both periodical $i$ and periodical $j$, the co-citation count between these two periodicals is incremented by one. This matrix thus captures the degree of relatedness between periodicals based on how frequently they are cited together in the literature. Similar to the citation matrix, we normalize the co-citation matrix by row, so that each entry reflects the relative tendency of a given periodical to be co-cited with others. The normalized rows of this matrix are then directly used as high-dimensional vector representations for each periodical.

The fourth (Co-citation+Node2Vec) constructs a periodical-periodical co-citation network based on the unnormalized co-citation matrix, where nodes represent periodicals and an edge is formed between two periodicals if they are ever co-cited by the same paper, with edge weights corresponding to their co-citation counts. We then apply Node2Vec on this co-citation network to learn low-dimensional vector embeddings for each periodical.

After obtaining periodical embeddings, we apply $k$-means to the periodical embeddings obtained to generate labels for the periodicals. The number of clusters was set to \(K = 26\), aligning with the 26 subject areas in the Scopus classification (excluding the multidisciplinary category). This choice facilitates direct comparison when evaluating clustering performance against Scopus labels.

\subsection{Document classification task}

The classification task aims to assess the coherence and interpretability of the labels generated by various schemes. To do that, we performed a document classification task using $23,322,430$ paper abstracts from the MAG. Table~S1 provides the field distribution of papers used in this work. The classifier employed was a combination of Hashing Vectorization~\cite{weinberger2009feature} for feature extraction and Complementary Naive Bayes (CNB)~\cite{rennie2003tackling} for classification. HashingVectorizer implements feature hashing for dimensionality reduction, mapping textual features to fixed-dimensional vectors through hash functions. Given a document $d$ containing tokens $(t_1, t_2, ..., t_n)$, the hashing transformation is defined as:
\begin{equation}
\mathbf{x}_d = \left[ \sum_{j:h(t_j)=1} f(t_j), \sum_{j:h(t_j)=2} f(t_j), ..., \sum_{j:h(t_j)=m} f(t_j) \right]
\end{equation}
where $h: \mathcal{T} \rightarrow \{1,2,...,m\}$ denotes the hash function mapping tokens to feature indices, $f(t_j)$ represents token frequency, and $m=2^{20}$ specifies the feature space dimensionality. This method eliminates vocabulary storage requirements while maintaining sparsity representation~\cite{weinberger2009feature}. CNB is a probabilistic classifier well-suited for imbalanced datasets. Different from the traditional Naive Bayes algorithm, which estimates the likelihood \(P(x_i | y)\) using data from class \(y\) itself, CNB estimates parameters based on the complement of each class—that is, all data \emph{not} belonging to class \(y\). This modification helps reduce bias towards majority classes and improves performance, especially in text classification tasks with skewed class distributions. For a document \(\mathbf{x} = (x_1, x_2, \ldots, x_n)\), CNB predicts the class \(y\) by maximizing the following score:
\[
\hat{y} = \arg\max_y \left[ \log P(y) - \sum_{i=1}^n x_i \log P(x_i \mid \bar{y}) \right]
\]
where \(P(y)\) is the prior probability of class \(y\), and \(P(x_i \mid \bar{y})\) is the likelihood of feature \(x_i\) given the complement of class \(y\).

We consider two settings for evaluating classification performance: multi-class classification and one-vs-rest classification. In the multi-class setting, the classification performance was evaluated using the macro and weighted mean of precision, recall, and F1-score across all classes as metrics. We further adopt two ranking-based metrics to evaluate classifier performance: Ranking Average Precision (RAP) and Ranking Loss. RAP measures how well, for each sample, the true label(s) are ranked above false labels, and is defined as
\begin{equation}
\text{RAP} = \frac{1}{N} \sum_{i=1}^{N} \frac{1}{|\mathcal{Y}_i|} \sum_{j \in \mathcal{Y}_i} \frac{|\{ k : f_i(j) \geq f_i(k),\ k \in \mathcal{Y}_i \}|}{r_{ij}}
\end{equation}
where $N$ is the number of samples, $\mathcal{Y}_i$ is the set of true labels for sample $i$, $f_i(j)$ is the predicted score for label $j$, and $r_{ij}$ is the rank of label $j$ for sample $i$. Ranking Loss quantifies the average fraction of incorrectly ordered label pairs (false label ranked above true label), and is computed as
\begin{equation}
\text{Ranking Loss} = \frac{1}{N} \sum_{i=1}^{N} \frac{1}{|\mathcal{Y}_i||\overline{\mathcal{Y}_i}|} \left| \left\{ (j, k): f_i(j) \leq f_i(k),\ j \in \mathcal{Y}_i,\ k \notin \mathcal{Y}_i \right\} \right|
\end{equation}
where $\overline{\mathcal{Y}_i}$ is the set of false labels for sample $i$. These metrics provide insights into how well the clustering-derived labels align with the semantic structure of the dataset.

In the one-vs-rest setting, we train $K$ classifiers and plot, for each classifier, the precision-recall (PR) and Receiver Operating Characteristic (ROC) curves at various thresholds. To provide a single performance metric across classifiers, we compute the macro-average Average Precision (AP) for the PR plots:
$$\text{Macro-AP} = \frac{1}{K} \sum_{k=1}^K \text{AP}_k ,$$
where $\text{AP}_k$ is the AP for class $k$. For ROC curves, we calculate the Area Under the Curve (AUC), defined as
$$\text{Macro-AUC} = \frac{1}{K} \sum_{k=1}^K \text{AUC}_k ,$$
where $\text{AUC}_k$ is the area under the ROC curve. The two macro metrics ensure that all classes are equally weighted, providing a balanced evaluation of the classifier's performance across classes.

\subsection{Topic-label consistency analysis}

To evaluate the semantic coherence between the labels assigned by $\mathcal{C}$ and the underlying topical structure of documents, we perform Latent Dirichlet Allocation (LDA)~\cite{blei2003latent} on the corpus of 23,322,430 paper abstracts. First, we determine the optimal number of topics by evaluating topic coherence scores across different numbers of topics, ranging from 10 to 200 with a step size of 10, on a sampled subset of 1,166,122 (5\%) abstracts. The coherence score ($C_v$)~\cite{roder2015exploring} measures the semantic similarity between high-scoring words in each topic, providing a quantitative assessment of topic interpretability. We select $T = 30 $ topics, which maximizes the coherence score (Fig.~S8).

For each document $d$ in our corpus, LDA provides a probability distribution over $T$ topics $\theta_d = (\theta_{d_1}, \ldots, \theta_{dK})$, and $\theta_{dk}$ represents the proportion of document $d$ that belongs to topic $k$. We assign each document its dominant topic $t_d$: $t_d = argmax_k \theta_{dk}$. To quantify the consistency between the topic assignments and various classification schemes, we computed Normalized Mutual Information (NMI), Adjusted Rand Index (ARI), and Fowlkes-Mallows Index (FMI). These metrics measure the degree of agreement between two different ways of partitioning the same data, and higher values suggest better alignment between the discovered topical structure and the classification scheme, indicating more semantically meaningful categorization. NMI is computed as:
\begin{equation}
\text{NMI} (T,C) = \frac{MI(T,C)}{\sqrt{H(T)H(C)}}
\end{equation}
where \(H(T)\) and \(H(C)\) are the entropy of topic assignments and class labels, respectively:
\begin{equation}
H(T) = -\sum_{t \in T} p(t) \log p(t)
\end{equation}
NMI falls between 0 and 1, where 1 indicates perfect correlation between topics and labels, and 0 indicates complete independence. ARI is defined as:
\begin{equation}
ARI = \frac{RI - E[RI]}{\max(RI) - E[RI]}, 
\end{equation}
where \(RI\) is the Rand Index, which measures the similarity between two types of data clustering, and \(E[RI]\) is the expected value of the Rand Index for a random clustering. Finally, FMI is computed as:
\begin{equation}
FMI = \sqrt{\frac{TP}{TP + FP} \cdot \frac{TP}{TP + FN}}, 
\end{equation}
where \(TP\) is the number of true positives, \(FP\) is the number of false positives, and \(FN\) is the number of false negatives.

\begin{acknowledgements}
This work was supported by the National Natural Science Foundation of China (72204206), City University of Hong Kong (Project No. 9610552, 7005968), and the Hong Kong Institute for Data Science.
\end{acknowledgements}

\bibliography{main}

\begin{thebibliography}{35}%
\makeatletter
\providecommand \@ifxundefined [1]{%
 \@ifx{#1\undefined}
}%
\providecommand \@ifnum [1]{%
 \ifnum #1\expandafter \@firstoftwo
 \else \expandafter \@secondoftwo
 \fi
}%
\providecommand \@ifx [1]{%
 \ifx #1\expandafter \@firstoftwo
 \else \expandafter \@secondoftwo
 \fi
}%
\providecommand \natexlab [1]{#1}%
\providecommand \enquote  [1]{``#1''}%
\providecommand \bibnamefont  [1]{#1}%
\providecommand \bibfnamefont [1]{#1}%
\providecommand \citenamefont [1]{#1}%
\providecommand \href@noop [0]{\@secondoftwo}%
\providecommand \href [0]{\begingroup \@sanitize@url \@href}%
\providecommand \@href[1]{\@@startlink{#1}\@@href}%
\providecommand \@@href[1]{\endgroup#1\@@endlink}%
\providecommand \@sanitize@url [0]{\catcode `\\12\catcode `\$12\catcode `\&12\catcode `\#12\catcode `\^12\catcode `\_12\catcode `\%12\relax}%
\providecommand \@@startlink[1]{}%
\providecommand \@@endlink[0]{}%
\providecommand \url  [0]{\begingroup\@sanitize@url \@url }%
\providecommand \@url [1]{\endgroup\@href {#1}{\urlprefix }}%
\providecommand \urlprefix  [0]{URL }%
\providecommand \Eprint [0]{\href }%
\providecommand \doibase [0]{https://doi.org/}%
\providecommand \selectlanguage [0]{\@gobble}%
\providecommand \bibinfo  [0]{\@secondoftwo}%
\providecommand \bibfield  [0]{\@secondoftwo}%
\providecommand \translation [1]{[#1]}%
\providecommand \BibitemOpen [0]{}%
\providecommand \bibitemStop [0]{}%
\providecommand \bibitemNoStop [0]{.\EOS\space}%
\providecommand \EOS [0]{\spacefactor3000\relax}%
\providecommand \BibitemShut  [1]{\csname bibitem#1\endcsname}%
\let\auto@bib@innerbib\@empty
\bibitem [{\citenamefont {Gl{\"a}nzel}\ and\ \citenamefont {Schubert}(2003)}]{glanzel2003new}%
  \BibitemOpen
  \bibfield  {author} {\bibinfo {author} {\bibfnamefont {W.}~\bibnamefont {Gl{\"a}nzel}}\ and\ \bibinfo {author} {\bibfnamefont {A.}~\bibnamefont {Schubert}},\ }\bibfield  {title} {\bibinfo {title} {A new classification scheme of science fields and subfields designed for scientometric evaluation purposes},\ }\href {https://doi.org/10.1023/A:1022378804087} {\bibfield  {journal} {\bibinfo  {journal} {Scientometrics}\ }\textbf {\bibinfo {volume} {56}},\ \bibinfo {pages} {357} (\bibinfo {year} {2003})}\BibitemShut {NoStop}%
\bibitem [{\citenamefont {Waltman}\ and\ \citenamefont {Van~Eck}(2012)}]{waltman2012new}%
  \BibitemOpen
  \bibfield  {author} {\bibinfo {author} {\bibfnamefont {L.}~\bibnamefont {Waltman}}\ and\ \bibinfo {author} {\bibfnamefont {N.~J.}\ \bibnamefont {Van~Eck}},\ }\bibfield  {title} {\bibinfo {title} {A new methodology for constructing a publication-level classification system of science},\ }\href {https://doi.org/10.1002/asi.22748} {\bibfield  {journal} {\bibinfo  {journal} {Journal of the American Society for Information Science and Technology}\ }\textbf {\bibinfo {volume} {63}},\ \bibinfo {pages} {2378} (\bibinfo {year} {2012})}\BibitemShut {NoStop}%
\bibitem [{\citenamefont {Zhang}\ \emph {et~al.}(2025)\citenamefont {Zhang}, \citenamefont {Gruber},\ and\ \citenamefont {Frietsch}}]{zhang2025impact}%
  \BibitemOpen
  \bibfield  {author} {\bibinfo {author} {\bibfnamefont {J.}~\bibnamefont {Zhang}}, \bibinfo {author} {\bibfnamefont {S.}~\bibnamefont {Gruber}},\ and\ \bibinfo {author} {\bibfnamefont {R.}~\bibnamefont {Frietsch}},\ }\bibfield  {title} {\bibinfo {title} {Impact of classification granularity on interdisciplinary performance assessment of research institutes and organizations},\ }\href {https://doi.org/10.2478/jdis-2025-0028} {\bibfield  {journal} {\bibinfo  {journal} {Journal of Data and Information Science}\ }\textbf {\bibinfo {volume} {10}},\ \bibinfo {pages} {61} (\bibinfo {year} {2025})}\BibitemShut {NoStop}%
\bibitem [{\citenamefont {Rafols}\ and\ \citenamefont {Leydesdorff}(2009)}]{rafols2009content}%
  \BibitemOpen
  \bibfield  {author} {\bibinfo {author} {\bibfnamefont {I.}~\bibnamefont {Rafols}}\ and\ \bibinfo {author} {\bibfnamefont {L.}~\bibnamefont {Leydesdorff}},\ }\bibfield  {title} {\bibinfo {title} {Content-based and algorithmic classifications of journals: Perspectives on the dynamics of scientific communication and indexer effects},\ }\href {https://doi.org/10.1002/asi.21086} {\bibfield  {journal} {\bibinfo  {journal} {Journal of the American Society for Information Science and Technology}\ }\textbf {\bibinfo {volume} {60}},\ \bibinfo {pages} {1823} (\bibinfo {year} {2009})}\BibitemShut {NoStop}%
\bibitem [{\citenamefont {Wang}\ and\ \citenamefont {Waltman}(2016)}]{wang2016large}%
  \BibitemOpen
  \bibfield  {author} {\bibinfo {author} {\bibfnamefont {Q.}~\bibnamefont {Wang}}\ and\ \bibinfo {author} {\bibfnamefont {L.}~\bibnamefont {Waltman}},\ }\bibfield  {title} {\bibinfo {title} {Large-scale analysis of the accuracy of the journal classification systems of web of science and scopus},\ }\href {https://doi.org/10.1016/j.joi.2016.02.003} {\bibfield  {journal} {\bibinfo  {journal} {Journal of Informetrics}\ }\textbf {\bibinfo {volume} {10}},\ \bibinfo {pages} {347} (\bibinfo {year} {2016})}\BibitemShut {NoStop}%
\bibitem [{\citenamefont {Pudovkin}\ and\ \citenamefont {Garfield}(2002)}]{pudovkin2002algorithmic}%
  \BibitemOpen
  \bibfield  {author} {\bibinfo {author} {\bibfnamefont {A.~I.}\ \bibnamefont {Pudovkin}}\ and\ \bibinfo {author} {\bibfnamefont {E.}~\bibnamefont {Garfield}},\ }\bibfield  {title} {\bibinfo {title} {Algorithmic procedure for finding semantically related journals},\ }\href {https://doi.org/10.1002/asi.10153} {\bibfield  {journal} {\bibinfo  {journal} {Journal of the American Society for Information Science and Technology}\ }\textbf {\bibinfo {volume} {53}},\ \bibinfo {pages} {1113} (\bibinfo {year} {2002})}\BibitemShut {NoStop}%
\bibitem [{\citenamefont {Leydesdorff}\ and\ \citenamefont {Bornmann}(2016)}]{leydesdorff2016operationalization}%
  \BibitemOpen
  \bibfield  {author} {\bibinfo {author} {\bibfnamefont {L.}~\bibnamefont {Leydesdorff}}\ and\ \bibinfo {author} {\bibfnamefont {L.}~\bibnamefont {Bornmann}},\ }\bibfield  {title} {\bibinfo {title} {The operationalization of “fields” as wos subject categories (wcs) in evaluative bibliometrics: The cases of “library and information science” and “science \& technology studies”},\ }\href {https://doi.org/10.1002/asi.23408} {\bibfield  {journal} {\bibinfo  {journal} {Journal of the Association for Information Science and Technology}\ }\textbf {\bibinfo {volume} {67}},\ \bibinfo {pages} {707} (\bibinfo {year} {2016})}\BibitemShut {NoStop}%
\bibitem [{\citenamefont {Thelwall}\ and\ \citenamefont {Pinfield}(2024)}]{thelwall2024accuracy}%
  \BibitemOpen
  \bibfield  {author} {\bibinfo {author} {\bibfnamefont {M.}~\bibnamefont {Thelwall}}\ and\ \bibinfo {author} {\bibfnamefont {S.}~\bibnamefont {Pinfield}},\ }\bibfield  {title} {\bibinfo {title} {The accuracy of field classifications for journals in scopus},\ }\href {https://doi.org/10.1007/s11192-023-04901-4} {\bibfield  {journal} {\bibinfo  {journal} {Scientometrics}\ }\textbf {\bibinfo {volume} {129}},\ \bibinfo {pages} {1097} (\bibinfo {year} {2024})}\BibitemShut {NoStop}%
\bibitem [{\citenamefont {Shen}\ \emph {et~al.}(2024)\citenamefont {Shen}, \citenamefont {Zhang},\ and\ \citenamefont {Zeng}}]{shen2024under}%
  \BibitemOpen
  \bibfield  {author} {\bibinfo {author} {\bibfnamefont {Z.}~\bibnamefont {Shen}}, \bibinfo {author} {\bibfnamefont {J.}~\bibnamefont {Zhang}},\ and\ \bibinfo {author} {\bibfnamefont {A.}~\bibnamefont {Zeng}},\ }\bibfield  {title} {\bibinfo {title} {Under-representativeness of physical chemistry journals},\ }\href@noop {} {\bibfield  {journal} {\bibinfo  {journal} {ChemRxiv}\ } (\bibinfo {year} {2024})}\BibitemShut {NoStop}%
\bibitem [{\citenamefont {Liao}\ \emph {et~al.}(2025)\citenamefont {Liao}, \citenamefont {Li},\ and\ \citenamefont {Shen}}]{liao2025journal}%
  \BibitemOpen
  \bibfield  {author} {\bibinfo {author} {\bibfnamefont {Y.}~\bibnamefont {Liao}}, \bibinfo {author} {\bibfnamefont {L.}~\bibnamefont {Li}},\ and\ \bibinfo {author} {\bibfnamefont {Z.}~\bibnamefont {Shen}},\ }\bibfield  {title} {\bibinfo {title} {Is journal citation indicator a good metric for art \& humanities journals currently?},\ }\href@noop {} {\bibfield  {journal} {\bibinfo  {journal} {arXiv preprint arXiv:2503.23400}\ } (\bibinfo {year} {2025})}\BibitemShut {NoStop}%
\bibitem [{\citenamefont {Janssens}\ \emph {et~al.}(2009)\citenamefont {Janssens}, \citenamefont {Zhang}, \citenamefont {De~Moor},\ and\ \citenamefont {Gl{\"a}nzel}}]{janssens2009hybrid}%
  \BibitemOpen
  \bibfield  {author} {\bibinfo {author} {\bibfnamefont {F.}~\bibnamefont {Janssens}}, \bibinfo {author} {\bibfnamefont {L.}~\bibnamefont {Zhang}}, \bibinfo {author} {\bibfnamefont {B.}~\bibnamefont {De~Moor}},\ and\ \bibinfo {author} {\bibfnamefont {W.}~\bibnamefont {Gl{\"a}nzel}},\ }\bibfield  {title} {\bibinfo {title} {Hybrid clustering for validation and improvement of subject-classification schemes},\ }\href {https://doi.org/10.1016/j.ipm.2009.06.003} {\bibfield  {journal} {\bibinfo  {journal} {Information Processing \& Management}\ }\textbf {\bibinfo {volume} {45}},\ \bibinfo {pages} {683} (\bibinfo {year} {2009})}\BibitemShut {NoStop}%
\bibitem [{\citenamefont {Thijs}\ \emph {et~al.}(2015)\citenamefont {Thijs}, \citenamefont {Zhang},\ and\ \citenamefont {Gl{\"a}nzel}}]{thijs2015bibliographic}%
  \BibitemOpen
  \bibfield  {author} {\bibinfo {author} {\bibfnamefont {B.}~\bibnamefont {Thijs}}, \bibinfo {author} {\bibfnamefont {L.}~\bibnamefont {Zhang}},\ and\ \bibinfo {author} {\bibfnamefont {W.}~\bibnamefont {Gl{\"a}nzel}},\ }\bibfield  {title} {\bibinfo {title} {Bibliographic coupling and hierarchical clustering for the validation and improvement of subject-classification schemes},\ }\href {https://doi.org/10.1007/s11192-015-1641-3} {\bibfield  {journal} {\bibinfo  {journal} {Scientometrics}\ }\textbf {\bibinfo {volume} {105}},\ \bibinfo {pages} {1453} (\bibinfo {year} {2015})}\BibitemShut {NoStop}%
\bibitem [{\citenamefont {Zhang}\ \emph {et~al.}(2010)\citenamefont {Zhang}, \citenamefont {Janssens}, \citenamefont {Liang},\ and\ \citenamefont {Gl{\"a}nzel}}]{zhang2010journal}%
  \BibitemOpen
  \bibfield  {author} {\bibinfo {author} {\bibfnamefont {L.}~\bibnamefont {Zhang}}, \bibinfo {author} {\bibfnamefont {F.}~\bibnamefont {Janssens}}, \bibinfo {author} {\bibfnamefont {L.}~\bibnamefont {Liang}},\ and\ \bibinfo {author} {\bibfnamefont {W.}~\bibnamefont {Gl{\"a}nzel}},\ }\bibfield  {title} {\bibinfo {title} {Journal cross-citation analysis for validation and improvement of journal-based subject classification in bibliometric research},\ }\href {https://doi.org/10.1007/s11192-010-0180-1} {\bibfield  {journal} {\bibinfo  {journal} {Scientometrics}\ }\textbf {\bibinfo {volume} {82}},\ \bibinfo {pages} {687} (\bibinfo {year} {2010})}\BibitemShut {NoStop}%
\bibitem [{\citenamefont {G{\'o}mez-N{\'u}{\~n}ez}\ \emph {et~al.}(2011)\citenamefont {G{\'o}mez-N{\'u}{\~n}ez}, \citenamefont {Vargas-Quesada}, \citenamefont {de~Moya-Aneg{\'o}n},\ and\ \citenamefont {Gl{\"a}nzel}}]{gomez2011improving}%
  \BibitemOpen
  \bibfield  {author} {\bibinfo {author} {\bibfnamefont {A.~J.}\ \bibnamefont {G{\'o}mez-N{\'u}{\~n}ez}}, \bibinfo {author} {\bibfnamefont {B.}~\bibnamefont {Vargas-Quesada}}, \bibinfo {author} {\bibfnamefont {F.}~\bibnamefont {de~Moya-Aneg{\'o}n}},\ and\ \bibinfo {author} {\bibfnamefont {W.}~\bibnamefont {Gl{\"a}nzel}},\ }\bibfield  {title} {\bibinfo {title} {Improving scimago journal \& country rank (sjr) subject classification through reference analysis},\ }\href {https://doi.org/10.1007/s11192-011-0485-8} {\bibfield  {journal} {\bibinfo  {journal} {Scientometrics}\ }\textbf {\bibinfo {volume} {89}},\ \bibinfo {pages} {741} (\bibinfo {year} {2011})}\BibitemShut {NoStop}%
\bibitem [{\citenamefont {G{\'o}mez-N{\'u}{\~n}ez}\ \emph {et~al.}(2014)\citenamefont {G{\'o}mez-N{\'u}{\~n}ez}, \citenamefont {Batagelj}, \citenamefont {Vargas-Quesada}, \citenamefont {Moya-Anegon},\ and\ \citenamefont {Chinchilla-Rodr{\'\i}guez}}]{gomez2014optimizing}%
  \BibitemOpen
  \bibfield  {author} {\bibinfo {author} {\bibfnamefont {A.~J.}\ \bibnamefont {G{\'o}mez-N{\'u}{\~n}ez}}, \bibinfo {author} {\bibfnamefont {V.}~\bibnamefont {Batagelj}}, \bibinfo {author} {\bibfnamefont {B.}~\bibnamefont {Vargas-Quesada}}, \bibinfo {author} {\bibfnamefont {F.}~\bibnamefont {Moya-Anegon}},\ and\ \bibinfo {author} {\bibfnamefont {Z.}~\bibnamefont {Chinchilla-Rodr{\'\i}guez}},\ }\bibfield  {title} {\bibinfo {title} {Optimizing scimago journal \& country rank classification by community detection},\ }\href {https://doi.org/10.1016/j.joi.2014.01.011} {\bibfield  {journal} {\bibinfo  {journal} {Journal of Informetrics}\ }\textbf {\bibinfo {volume} {8}},\ \bibinfo {pages} {369} (\bibinfo {year} {2014})}\BibitemShut {NoStop}%
\bibitem [{\citenamefont {G{\'o}mez-N{\'u}{\~n}ez}\ \emph {et~al.}(2016)\citenamefont {G{\'o}mez-N{\'u}{\~n}ez}, \citenamefont {Vargas-Quesada},\ and\ \citenamefont {de~Moya-Aneg{\'o}n}}]{gomez2016updating}%
  \BibitemOpen
  \bibfield  {author} {\bibinfo {author} {\bibfnamefont {A.~J.}\ \bibnamefont {G{\'o}mez-N{\'u}{\~n}ez}}, \bibinfo {author} {\bibfnamefont {B.}~\bibnamefont {Vargas-Quesada}},\ and\ \bibinfo {author} {\bibfnamefont {F.}~\bibnamefont {de~Moya-Aneg{\'o}n}},\ }\bibfield  {title} {\bibinfo {title} {Updating the sci mago journal and country rank classification: A new approach using w ard's clustering and alternative combination of citation measures},\ }\href {https://doi.org/10.1002/asi.23370} {\bibfield  {journal} {\bibinfo  {journal} {Journal of the Association for Information Science and Technology}\ }\textbf {\bibinfo {volume} {67}},\ \bibinfo {pages} {178} (\bibinfo {year} {2016})}\BibitemShut {NoStop}%
\bibitem [{\citenamefont {Peng}\ \emph {et~al.}(2021)\citenamefont {Peng}, \citenamefont {Ke}, \citenamefont {Budak}, \citenamefont {Romero},\ and\ \citenamefont {Ahn}}]{peng2021neural}%
  \BibitemOpen
  \bibfield  {author} {\bibinfo {author} {\bibfnamefont {H.}~\bibnamefont {Peng}}, \bibinfo {author} {\bibfnamefont {Q.}~\bibnamefont {Ke}}, \bibinfo {author} {\bibfnamefont {C.}~\bibnamefont {Budak}}, \bibinfo {author} {\bibfnamefont {D.~M.}\ \bibnamefont {Romero}},\ and\ \bibinfo {author} {\bibfnamefont {Y.-Y.}\ \bibnamefont {Ahn}},\ }\bibfield  {title} {\bibinfo {title} {Neural embeddings of scholarly periodicals reveal complex disciplinary organizations},\ }\href {https://doi.org/10.1126/sciadv.abb9004} {\bibfield  {journal} {\bibinfo  {journal} {Science Advances}\ }\textbf {\bibinfo {volume} {7}},\ \bibinfo {pages} {eabb9004} (\bibinfo {year} {2021})}\BibitemShut {NoStop}%
\bibitem [{\citenamefont {Lyu}\ and\ \citenamefont {Ke}(2025)}]{lyu2025mapping}%
  \BibitemOpen
  \bibfield  {author} {\bibinfo {author} {\bibfnamefont {Z.}~\bibnamefont {Lyu}}\ and\ \bibinfo {author} {\bibfnamefont {Q.}~\bibnamefont {Ke}},\ }\bibfield  {title} {\bibinfo {title} {Mapping the changing structure of science through diachronic periodical embeddings},\ }\href {https://doi.org/https://doi.org/10.1016/j.chaos.2025.117295} {\bibfield  {journal} {\bibinfo  {journal} {Chaos, Solitons \& Fractals}\ }\textbf {\bibinfo {volume} {201}},\ \bibinfo {pages} {117295} (\bibinfo {year} {2025})}\BibitemShut {NoStop}%
\bibitem [{\citenamefont {Mikolov}\ \emph {et~al.}(2013)\citenamefont {Mikolov}, \citenamefont {Sutskever}, \citenamefont {Chen}, \citenamefont {Corrado},\ and\ \citenamefont {Dean}}]{tomas2013sgns}%
  \BibitemOpen
  \bibfield  {author} {\bibinfo {author} {\bibfnamefont {T.}~\bibnamefont {Mikolov}}, \bibinfo {author} {\bibfnamefont {I.}~\bibnamefont {Sutskever}}, \bibinfo {author} {\bibfnamefont {K.}~\bibnamefont {Chen}}, \bibinfo {author} {\bibfnamefont {G.}~\bibnamefont {Corrado}},\ and\ \bibinfo {author} {\bibfnamefont {J.}~\bibnamefont {Dean}},\ }\bibfield  {title} {\bibinfo {title} {Distributed representations of words and phrases and their compositionality},\ }in\ \href@noop {} {\emph {\bibinfo {booktitle} {Proceedings of the 26th International Conference on Neural Information Processing Systems - Volume 2}}},\ \bibinfo {series and number} {NIPS'13}\ (\bibinfo  {publisher} {Curran Associates Inc.},\ \bibinfo {address} {Red Hook, NY, USA},\ \bibinfo {year} {2013})\ p.\ \bibinfo {pages} {3111–3119}\BibitemShut {NoStop}%
\bibitem [{\citenamefont {Boyack}\ \emph {et~al.}(2005)\citenamefont {Boyack}, \citenamefont {Klavans},\ and\ \citenamefont {B{\"o}rner}}]{boyack2005mapping}%
  \BibitemOpen
  \bibfield  {author} {\bibinfo {author} {\bibfnamefont {K.~W.}\ \bibnamefont {Boyack}}, \bibinfo {author} {\bibfnamefont {R.}~\bibnamefont {Klavans}},\ and\ \bibinfo {author} {\bibfnamefont {K.}~\bibnamefont {B{\"o}rner}},\ }\bibfield  {title} {\bibinfo {title} {Mapping the backbone of science},\ }\href {https://doi.org/10.1007/s11192-005-0255-6} {\bibfield  {journal} {\bibinfo  {journal} {Scientometrics}\ }\textbf {\bibinfo {volume} {64}},\ \bibinfo {pages} {351} (\bibinfo {year} {2005})}\BibitemShut {NoStop}%
\bibitem [{\citenamefont {Davis}\ and\ \citenamefont {Goadrich}(2006)}]{davis2006relationship}%
  \BibitemOpen
  \bibfield  {author} {\bibinfo {author} {\bibfnamefont {J.}~\bibnamefont {Davis}}\ and\ \bibinfo {author} {\bibfnamefont {M.}~\bibnamefont {Goadrich}},\ }\bibfield  {title} {\bibinfo {title} {The relationship between precision-recall and roc curves},\ }in\ \href {https://doi.org/10.1145/1143844.1143874} {\emph {\bibinfo {booktitle} {Proceedings of the 23rd International Conference on Machine learning}}}\ (\bibinfo {year} {2006})\ pp.\ \bibinfo {pages} {233--240}\BibitemShut {NoStop}%
\bibitem [{\citenamefont {Blei}\ \emph {et~al.}(2003)\citenamefont {Blei}, \citenamefont {Ng},\ and\ \citenamefont {Jordan}}]{blei2003latent}%
  \BibitemOpen
  \bibfield  {author} {\bibinfo {author} {\bibfnamefont {D.~M.}\ \bibnamefont {Blei}}, \bibinfo {author} {\bibfnamefont {A.~Y.}\ \bibnamefont {Ng}},\ and\ \bibinfo {author} {\bibfnamefont {M.~I.}\ \bibnamefont {Jordan}},\ }\bibfield  {title} {\bibinfo {title} {Latent dirichlet allocation},\ }\href@noop {} {\bibfield  {journal} {\bibinfo  {journal} {Journal of Machine Learning Research}\ }\textbf {\bibinfo {volume} {3}},\ \bibinfo {pages} {993} (\bibinfo {year} {2003})}\BibitemShut {NoStop}%
\bibitem [{\citenamefont {Maaten}\ and\ \citenamefont {Hinton}(2008)}]{van2008visualizing}%
  \BibitemOpen
  \bibfield  {author} {\bibinfo {author} {\bibfnamefont {L.~v.~d.}\ \bibnamefont {Maaten}}\ and\ \bibinfo {author} {\bibfnamefont {G.}~\bibnamefont {Hinton}},\ }\bibfield  {title} {\bibinfo {title} {Visualizing data using t-sne},\ }\href@noop {} {\bibfield  {journal} {\bibinfo  {journal} {Journal of Machine Learning Research}\ }\textbf {\bibinfo {volume} {9}},\ \bibinfo {pages} {2579} (\bibinfo {year} {2008})}\BibitemShut {NoStop}%
\bibitem [{\citenamefont {Gates}\ \emph {et~al.}(2019)\citenamefont {Gates}, \citenamefont {Wood}, \citenamefont {Hetrick},\ and\ \citenamefont {Ahn}}]{gates2019element}%
  \BibitemOpen
  \bibfield  {author} {\bibinfo {author} {\bibfnamefont {A.~J.}\ \bibnamefont {Gates}}, \bibinfo {author} {\bibfnamefont {I.~B.}\ \bibnamefont {Wood}}, \bibinfo {author} {\bibfnamefont {W.~P.}\ \bibnamefont {Hetrick}},\ and\ \bibinfo {author} {\bibfnamefont {Y.-Y.}\ \bibnamefont {Ahn}},\ }\bibfield  {title} {\bibinfo {title} {Element-centric clustering comparison unifies overlaps and hierarchy},\ }\href {https://doi.org/10.1038/s41598-019-44892-y} {\bibfield  {journal} {\bibinfo  {journal} {Scientific Reports}\ }\textbf {\bibinfo {volume} {9}},\ \bibinfo {pages} {8574} (\bibinfo {year} {2019})}\BibitemShut {NoStop}%
\bibitem [{\citenamefont {Shepard}(1968)}]{shepard1968two}%
  \BibitemOpen
  \bibfield  {author} {\bibinfo {author} {\bibfnamefont {D.}~\bibnamefont {Shepard}},\ }\bibfield  {title} {\bibinfo {title} {A two-dimensional interpolation function for irregularly-spaced data},\ }in\ \href {https://doi.org/10.1145/800186.81061} {\emph {\bibinfo {booktitle} {Proceedings of the 1968 23rd ACM national conference}}}\ (\bibinfo {year} {1968})\ pp.\ \bibinfo {pages} {517--524}\BibitemShut {NoStop}%
\bibitem [{\citenamefont {Xu}\ \emph {et~al.}(2023)\citenamefont {Xu}, \citenamefont {Zhu},\ and\ \citenamefont {Clifton}}]{xu2023multimodal}%
  \BibitemOpen
  \bibfield  {author} {\bibinfo {author} {\bibfnamefont {P.}~\bibnamefont {Xu}}, \bibinfo {author} {\bibfnamefont {X.}~\bibnamefont {Zhu}},\ and\ \bibinfo {author} {\bibfnamefont {D.~A.}\ \bibnamefont {Clifton}},\ }\bibfield  {title} {\bibinfo {title} {Multimodal learning with transformers: A survey},\ }\href {https://doi.org/10.1109/TPAMI.2023.3275156} {\bibfield  {journal} {\bibinfo  {journal} {IEEE Transactions on Pattern Analysis and Machine Intelligence}\ }\textbf {\bibinfo {volume} {45}},\ \bibinfo {pages} {12113} (\bibinfo {year} {2023})}\BibitemShut {NoStop}%
\bibitem [{\citenamefont {Beltagy}\ \emph {et~al.}(2019)\citenamefont {Beltagy}, \citenamefont {Lo},\ and\ \citenamefont {Cohan}}]{beltagy2019scibert}%
  \BibitemOpen
  \bibfield  {author} {\bibinfo {author} {\bibfnamefont {I.}~\bibnamefont {Beltagy}}, \bibinfo {author} {\bibfnamefont {K.}~\bibnamefont {Lo}},\ and\ \bibinfo {author} {\bibfnamefont {A.}~\bibnamefont {Cohan}},\ }\bibfield  {title} {\bibinfo {title} {Scibert: A pretrained language model for scientific text},\ }\bibfield  {journal} {\bibinfo  {journal} {arXiv preprint arXiv:1903.10676}\ }\href {https://doi.org/10.18653/v1/D19-1371} {10.18653/v1/D19-1371} (\bibinfo {year} {2019})\BibitemShut {NoStop}%
\bibitem [{\citenamefont {Dong}\ \emph {et~al.}(2024)\citenamefont {Dong}, \citenamefont {Lyu},\ and\ \citenamefont {Ke}}]{dong2024towards}%
  \BibitemOpen
  \bibfield  {author} {\bibinfo {author} {\bibfnamefont {J.}~\bibnamefont {Dong}}, \bibinfo {author} {\bibfnamefont {Z.}~\bibnamefont {Lyu}},\ and\ \bibinfo {author} {\bibfnamefont {Q.}~\bibnamefont {Ke}},\ }\bibfield  {title} {\bibinfo {title} {Towards understanding evolution of science through language model series},\ }\href@noop {} {\bibfield  {journal} {\bibinfo  {journal} {arXiv preprint arXiv:2409.09636}\ } (\bibinfo {year} {2024})}\BibitemShut {NoStop}%
\bibitem [{\citenamefont {Sinha}\ \emph {et~al.}(2015)\citenamefont {Sinha}, \citenamefont {Shen}, \citenamefont {Song}, \citenamefont {Ma}, \citenamefont {Eide}, \citenamefont {Hsu},\ and\ \citenamefont {Wang}}]{sinha2015overview}%
  \BibitemOpen
  \bibfield  {author} {\bibinfo {author} {\bibfnamefont {A.}~\bibnamefont {Sinha}}, \bibinfo {author} {\bibfnamefont {Z.}~\bibnamefont {Shen}}, \bibinfo {author} {\bibfnamefont {Y.}~\bibnamefont {Song}}, \bibinfo {author} {\bibfnamefont {H.}~\bibnamefont {Ma}}, \bibinfo {author} {\bibfnamefont {D.}~\bibnamefont {Eide}}, \bibinfo {author} {\bibfnamefont {B.-j.~P.}\ \bibnamefont {Hsu}},\ and\ \bibinfo {author} {\bibfnamefont {K.}~\bibnamefont {Wang}},\ }\bibfield  {title} {\bibinfo {title} {An overview of microsoft academic service (mas) and applications},\ }in\ \href {https://doi.org/10.1145/2740908.2742839} {\emph {\bibinfo {booktitle} {Proceedings of the 24th International Conference on World Wide Web}}}\ (\bibinfo {year} {2015})\ p.\ \bibinfo {pages} {243–246}\BibitemShut {NoStop}%
\bibitem [{\citenamefont {Elsevier}()}]{elsevier-no-date-scopus-content}%
  \BibitemOpen
  \bibfield  {author} {\bibinfo {author} {\bibnamefont {Elsevier}},\ }\href {https://www.elsevier.com/solutions/scopus/how-scopus-works/content} {\bibinfo {title} {{Content - How Scopus Works - Scopus - | Elsevier solutions}}}\BibitemShut {NoStop}%
\bibitem [{\citenamefont {Grover}\ and\ \citenamefont {Leskovec}(2016)}]{grover2016node2vec}%
  \BibitemOpen
  \bibfield  {author} {\bibinfo {author} {\bibfnamefont {A.}~\bibnamefont {Grover}}\ and\ \bibinfo {author} {\bibfnamefont {J.}~\bibnamefont {Leskovec}},\ }\bibfield  {title} {\bibinfo {title} {node2vec: Scalable feature learning for networks},\ }in\ \href {https://doi.org/10.1145/2939672.2939754} {\emph {\bibinfo {booktitle} {Proceedings of the 22nd ACM SIGKDD International Conference on Knowledge Discovery and Data Mining}}}\ (\bibinfo {year} {2016})\ pp.\ \bibinfo {pages} {855--864}\BibitemShut {NoStop}%
\bibitem [{\citenamefont {Leskovec}\ and\ \citenamefont {Sosi{\v c}}(2016)}]{leskovecSNAPGeneralPurposeNetwork2016}%
  \BibitemOpen
  \bibfield  {author} {\bibinfo {author} {\bibfnamefont {J.}~\bibnamefont {Leskovec}}\ and\ \bibinfo {author} {\bibfnamefont {R.}~\bibnamefont {Sosi{\v c}}},\ }\bibfield  {title} {\bibinfo {title} {{{SNAP}}: {{A General-Purpose Network Analysis}} and {{Graph-Mining Library}}},\ }\href {https://doi.org/10.1145/2898361} {\bibfield  {journal} {\bibinfo  {journal} {ACM Trans. Intell. Syst. Technol.}\ }\textbf {\bibinfo {volume} {8}},\ \bibinfo {pages} {1:1} (\bibinfo {year} {2016})}\BibitemShut {NoStop}%
\bibitem [{\citenamefont {Weinberger}\ \emph {et~al.}(2009)\citenamefont {Weinberger}, \citenamefont {Dasgupta}, \citenamefont {Langford}, \citenamefont {Smola},\ and\ \citenamefont {Attenberg}}]{weinberger2009feature}%
  \BibitemOpen
  \bibfield  {author} {\bibinfo {author} {\bibfnamefont {K.}~\bibnamefont {Weinberger}}, \bibinfo {author} {\bibfnamefont {A.}~\bibnamefont {Dasgupta}}, \bibinfo {author} {\bibfnamefont {J.}~\bibnamefont {Langford}}, \bibinfo {author} {\bibfnamefont {A.}~\bibnamefont {Smola}},\ and\ \bibinfo {author} {\bibfnamefont {J.}~\bibnamefont {Attenberg}},\ }\bibfield  {title} {\bibinfo {title} {Feature hashing for large scale multitask learning},\ }in\ \href {https://doi.org/10.1145/1553374.1553516} {\emph {\bibinfo {booktitle} {ICML}}}\ (\bibinfo {year} {2009})\ pp.\ \bibinfo {pages} {1113--1120}\BibitemShut {NoStop}%
\bibitem [{\citenamefont {Rennie}\ \emph {et~al.}(2003)\citenamefont {Rennie}, \citenamefont {Shih}, \citenamefont {Teevan},\ and\ \citenamefont {Karger}}]{rennie2003tackling}%
  \BibitemOpen
  \bibfield  {author} {\bibinfo {author} {\bibfnamefont {J.~D.}\ \bibnamefont {Rennie}}, \bibinfo {author} {\bibfnamefont {L.}~\bibnamefont {Shih}}, \bibinfo {author} {\bibfnamefont {J.}~\bibnamefont {Teevan}},\ and\ \bibinfo {author} {\bibfnamefont {D.~R.}\ \bibnamefont {Karger}},\ }\bibfield  {title} {\bibinfo {title} {Tackling the poor assumptions of naive bayes text classifiers},\ }in\ \href@noop {} {\emph {\bibinfo {booktitle} {Proceedings of the 20th International Conference on Machine Learning (ICML-03)}}}\ (\bibinfo {year} {2003})\ pp.\ \bibinfo {pages} {616--623}\BibitemShut {NoStop}%
\bibitem [{\citenamefont {R{\"o}der}\ \emph {et~al.}(2015)\citenamefont {R{\"o}der}, \citenamefont {Both},\ and\ \citenamefont {Hinneburg}}]{roder2015exploring}%
  \BibitemOpen
  \bibfield  {author} {\bibinfo {author} {\bibfnamefont {M.}~\bibnamefont {R{\"o}der}}, \bibinfo {author} {\bibfnamefont {A.}~\bibnamefont {Both}},\ and\ \bibinfo {author} {\bibfnamefont {A.}~\bibnamefont {Hinneburg}},\ }\bibfield  {title} {\bibinfo {title} {Exploring the space of topic coherence measures},\ }in\ \href {https://doi.org/10.1145/2684822.268532} {\emph {\bibinfo {booktitle} {Proceedings of the eighth ACM International Conference on Web Search and Data Mining}}}\ (\bibinfo {year} {2015})\ pp.\ \bibinfo {pages} {399--408}\BibitemShut {NoStop}%
\end{thebibliography}%
\end{document}